\documentclass[%
reprint,
superscriptaddress,
 amsmath,amssymb,
 aps,
prl,
]{revtex4-2}

\usepackage{graphicx}
\usepackage{dcolumn}
\usepackage{bm}
\usepackage{float}

\begin{document}


\title{The electric-field-driven intermediate state of three-dimensional superconductors}

\author{Ion Cojocari}
\affiliation{Universit\'{e} Paris-Saclay, CNRS/IN2P3, IJCLab, 91405 Orsay, France}

\author{Enzo Andreani}
\affiliation{Universit\'e Paris-Saclay, CNRS, Laboratoire de Physique des Solides, 91405, Orsay, France}

\author{Paola Verni\`{e}re}
\affiliation{Universit\'{e} Paris-Saclay, CNRS/IN2P3, IJCLab, 91405 Orsay, France}

\author{Florian Pallier}
\affiliation{Universit\'{e} Paris-Saclay, CNRS/IN2P3, IJCLab, 91405 Orsay, France}

\author{Marc Gabay}
\affiliation{Universit\'e Paris-Saclay, CNRS, Laboratoire de Physique des Solides, 91405, Orsay, France}

\author{Miguel Monteverde}
\affiliation{Universit\'e Paris-Saclay, CNRS, Laboratoire de Physique des Solides, 91405, Orsay, France}

\author{Claire Marrache-Kikuchi}
\affiliation{Universit\'{e} Paris-Saclay, CNRS/IN2P3, IJCLab, 91405 Orsay, France}

\author{Shamashis Sengupta}
\email[]{shamashis.sengupta@ijclab.in2p3.fr}
\affiliation{Universit\'{e} Paris-Saclay, CNRS/IN2P3, IJCLab, 91405 Orsay, France}

\begin{abstract}

The coexistence of superconductivity and finite electric fields may enable access to intriguing forms of electronic states. We demonstrate the emergence of an intermediate state in which electric fields penetrate the system while superconductivity still persists. Our measurements reveal a nonclassical regime characterized by the simultaneous presence of supercurrent and dissipative charge transport. This state, realized in a pristine unpatterned three-dimensional system, arises from electric-field-driven order parameter fluctuations. It provides a platform to explore dissipative states of charged quantum fluids far from equilibrium.

\end{abstract}

\maketitle

\newpage




\newpage

The earliest comprehensive phenomenological description of superconductivity was provided by the London theory\cite{london1,london2,tinkham} published in 1935. According to the first London equation, an electric field $\mathbf{E}$ is related to the variation of supercurrent density $\mathbf{j_s}$ with time $t$: $\mathbf{E}\propto\frac{\partial \mathbf{j_s}}{\partial t}$. It implies that if a steady electric field is present inside a superconductor, the supercurrent diverges to infinity without any bound as time increases. This situation is unphysical. Yet it poses a challenging question - what property does a superconductor actually exhibit when an electric field is applied? This issue has received attention in specific cases of low-dimensional devices where dissipation arises from phase slips\cite{tinkham,skocpol,chen,arutyunov,ivlev,lindelof}, and which has been relevant for applications in single-photon detection\cite{goltsman}. But the more general case of electric fields in a macroscale three-dimensional (3D) system has rarely been addressed. Here we report the outcome of applying voltage gradients across large superconducting films, leading to the observation of nonequilibrium states driven by order parameter fluctuations. The results reveal that a zero-resistance supercurrent can coexist with dissipative charge flow. This is a nonclassical mode of transport which highlights that electric field penetration does not necessarily imply a complete disappearance of superconductivity. Our work demonstrates that it is possible to conceive of an electric-field-driven intermediate state (EFDIS) in superconductors that is still waiting to be properly investigated.

In Ginzburg-Landau theory and Bardeen-Cooper-Schrieffer (BCS) theory, the superconducting state is described by a complex order parameter. Incorporating an electric field in these theories poses a challenging problem, as it implies dealing with a nonequilibrium state. An important development in this context was the discovery of the Josephson effect\cite{tinkham,josephson}, which provides the platform for realizing quantum electronic circuits\cite{nakamura,vion}, radiation emitters\cite{welp}, and voltage standards\cite{hamilton}. It refers to the tunnelling of Cooper pairs across a weak link (having length $L$) smaller than the coherence length $\xi$ of superconducting leads. If the phase difference on two sides of the weak link is $\varphi$, then the voltage $v$ is given by $v=\frac{\hbar}{2e}\frac{d\varphi}{dt}$. This provides the basis of our understanding that voltages inside superconducting systems are related to time variations of the complex order parameter. The Josephson effect in its original form is constrained by the criteria that there has to be some structure in the form of a weak link within the superconducting system, and at least one of the dimensions should be sufficiently small and comparable to the coherence length. Here we are concerned with a different problem in the opposite limit - how does a pristine unpatterned 3D superconductor of arbitrary length respond to an electric field? We describe next our experiments to obtain insights about this issue.

\begin{figure*}
\begin{center}
\includegraphics[width=176mm]{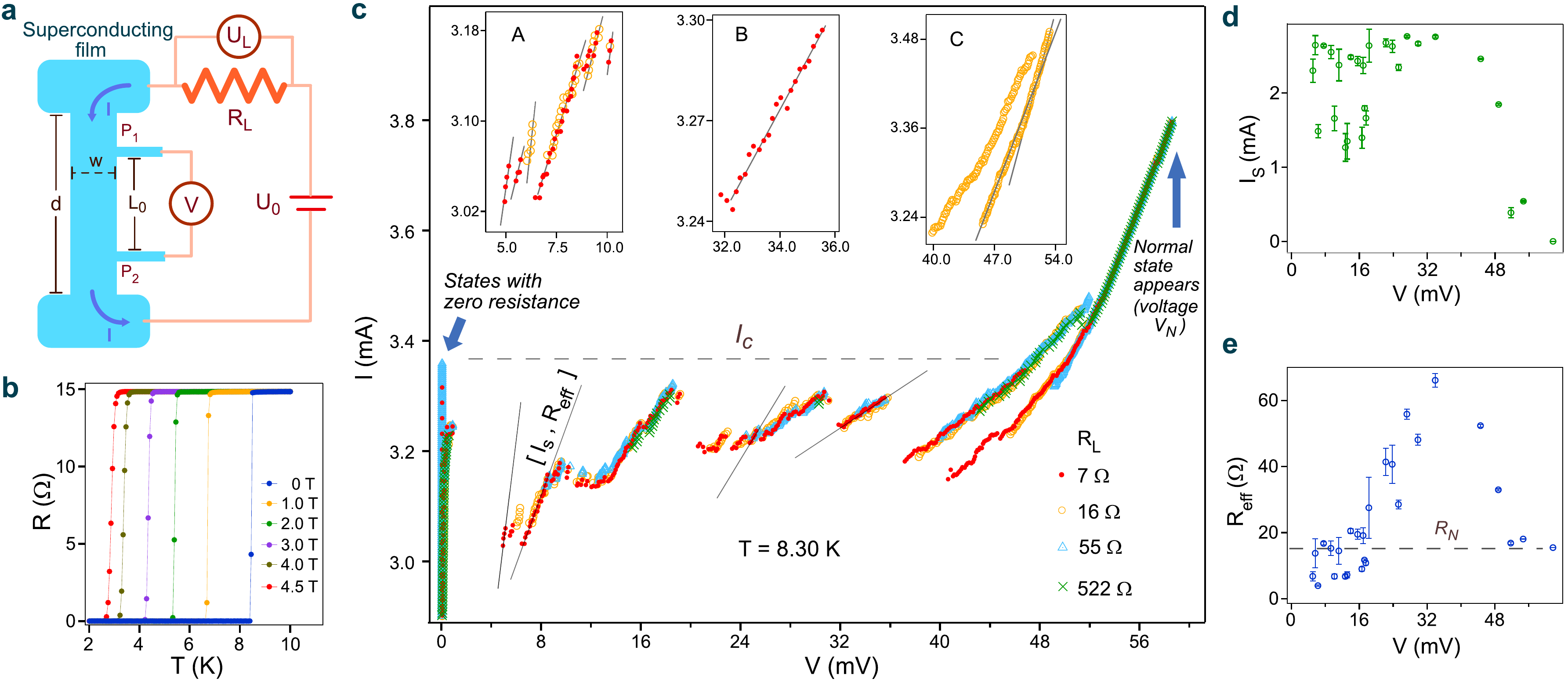}
\caption{\textbf{(a)} Circuit implemented for measuring transport behaviour of the superconducting film without current-biasing the sample. \textbf{(b)} Resistance ($R$) of the film measured in four-probe configuration as a function of temperature ($T$), with magnetic fields applied in the perpendicular direction. \textbf{(c)} $I$-$V$ plots obtained by increasing the dc voltage $U_0$. Insets A, B, C: Close-up view of data points in some selected regions. \textbf{(d,e)} Values of the intercept on current-axis ($I_S$) and effective resistance ($R_{eff}$), extracted from linear fits at different parts of the $I$-$V$ curves, plotted as a function of $V$. Each point corresponds to the mean value of the voltage range used for the linear fit.}
\end{center}
\end{figure*}

Our experiments were conducted on Nb films in four-probe geometry (Fig. 1a). The total length ($d$) and width ($w$) of the central channel are 960 $\mu$m and 85 $\mu$m respectively. The length $L_0$ between voltage probes (marked $P_1$ and $P_2$ in Fig. 1a) is 584 $\mu$m. The thickness ($h$) is 85 nm. Fig. 1b shows the measured four-probe resistance ($R$) as a function of temperature ($T$) for different values of applied magnetic field ($H$) in the perpendicular direction. The critical temperature ($T_c$) at zero field is 8.45 K. The Ginzburg-Landau coherence length is estimated from critical magnetic field measurements to be 10.1 nm. It is significantly shorter than all the dimensions of the film confirming its 3D nature. The four-probe resistance in the normal state ($R_N$) is 14.9 $\Omega$.

For probing current-voltage relations (Fig. 1a), a dc voltage $U_0$ is applied to the series combination of the superconducting film and an external resistance $R_L$. The latter is placed outside the cryostat at room temperature. $R_L$ is chosen to be small so that the Nb film is not current-biased upon developing finite resistance\cite{sengupta}. The four-probe voltage $V$ is measured across the superconductor as $U_0$ is ramped up\cite{supp}. The current ($I$) is monitored by measuring the voltage $U_L$ across the resistor $R_L$. These measurements allow us to determine the $I$-$V$ curve in the form of a scatter plot. In Fig. 1c, we show a set of $I$-$V$ plots with the superconductor maintained at $T$ = 8.30 K. The measurement was repeated using four different values of $R_L$. Initially, as $U_0$ is increased, the superconductor remains in the zero-resistance state till a switching current $I_c$ is reached (at $U_0$=$I_cR_L$)\cite{supp}. Beyond this point, finite $V$ develops across the superconductor indicating the onset of dissipation and the presence of electric fields inside. $I$ shows variations within a small range of values (between 3.0 - 3.4 mA) as $V$ increases. The completely normal state appears at a voltage $V_N \approx$ 60 mV, where the slope $\frac{dI}{dV}$ approaches $R_N^{-1}$. The range of $V$ below this value represents an electric-field-driven intermediate state (EFDIS) of the superconductor.

There are several noteworthy aspects about the distribution of data points in Fig. 1c. Each point having a finite value of $V$ represents a nonequilibrium state of the system. The points corresponding to measurements with four different $R_L$ values mostly superpose on each other and trace out specific patterns. In many places, consecutive data points arrange themselves along straight lines. A few of these linear segments are indicated in Fig. 1c. This allows us to determine effective values of conductance ($g_{eff}$) and resistance ($R_{eff}$) from the slope, representing a property of dissipative charge carriers. $g_{eff}$=$R_{eff}^{-1}$=$\frac{dI}{dV}$. Unlike a usual metal, the straight lines do not extrapolate to pass through the origin. Instead, they have a finite intercept $I_S$ on the current-axis. The intercept $I_S$ can be interpreted as a current  without a voltage drop, i.e., a supercurrent. The voltage $V$ corresponds to a dissipative current component $I_R=V/R_{eff}$. The expression for the total current is $I$=$I_S$+$I_R$. Along each of the linear segments (Fig. 1c) in the EFDIS, the \emph{macrostate} of the system can be described by the set of two quantities [$I_S$,$R_{eff}$]. These were estimated from the $I$-$V$ plots where sufficient data points (at least three) were present to allow a linear fit. Some examples of linear fits are shown in the insets of Fig. 1c. The estimates of $I_S$ and $R_{eff}$ are shown in Figs. 1d and 1e. $I_S$ reduces to zero at large values of $V$ (Fig. 1d), which is expected since large voltages completely destroy superconductivity. The observed values of $R_{eff}$ (Fig. 1e) underscore a very interesting phenomenon which will be discussed later.

A key factor that allowed us to observe the nonequilibrium intermediate state featuring both a supercurrent and a voltage drop is that the sample was not current-biased. In Fig. 1c, a large value of 522 $\Omega$ for $R_L$  is comparatively much closer to current-biasing conditions than the lowest value of 7 $\Omega$. The transition tends to be sharper for $R_L$ = 522 $\Omega$ with fewer data points along the $V$-axis. On reducing $R_L$, we were able to uncover much finer details of nonequilibrium states on the $I$-$V$ plane. This is because lowering of $R_L$ compels a larger fraction of the applied voltage $U_0$ to drop across the superconductor.

The experiments described above were also carried out at a lower temperature of 2.0 K and with an applied magnetic field of 3.0 T. The result of the $I$-$V$ measurement (with $R_L$ = 6 $\Omega$) is presented in Fig. 2a. Here extremely clear linear slopes were observed (marked by an ellipse). $I_S$ and $R_{eff}$ extracted from this plot are displayed in Figs. 2b and 2c. In Fig. 2a, the sharp linear segments all lie at current values lower than the critical current $I_c$ resulting in a non-monotonic $I$-$V$ curve. Current-biased measurements would not be able to access these nonequilibrium states.

The presence of linear segments in the $I$-$V$ curve with finite $I$-axis intercept is a key aspect of the EFDIS. It implies that the total current can be decomposed into a supercurrent component and a resistive component. In classical circuit theory, the addition of two currents implies the existence of two conducting pathways \emph{in parallel}. We may hypothesize that there is a spatial segregation of superconducting and normal phases - which allows the propagation of these two physically separated conduction channels. However, this violates Kirchhoff's voltage law (KVL) since the voltage drop along the two paths would be different - zero for one and finite for the other. Therefore, the charge current in the EFDIS requires a nonclassical description. An analogous situation is known\cite{tinkham,lindelof,timm} for resistively-shunted Josephson junctions (JJs), in which case a supercurrent and a dissipative current can propagate simultaneously due to dynamic order parameter fluctuations through a superconducting weak link. The model of a network of JJs has been applied in the past to explain the characteristics of quasi-2D superconductors (e.g. in SrTiO$_3$-based heterostructures\cite{hurand}), where superconducting islands are separated by non-superconducting weak link regions. Such systems typically show broad transitions in resistivity measurements due to effects of disorder and non-uniform distribution of local $T_c$. In contrast, our films show very sharp transitions (Fig. 1b) and a well-defined $T_c$ of 8.45 K, which is consistent with thick metallic Nb films. There is no evident reason to model our system as a network of JJs, even though the interrelationship of order parameter fluctuations and finite voltage gradients needs be included in a heuristic model for explaining our results. Based on these considerations, we proceed to carry out an analysis using the framework of Ginzburg-Landau theory. It will be applicable for arbitrarily long samples ($L\!>>\!\xi$), to highlight that we are dealing with uniform superconducting films rather than a network of superconducting islands coupled by the Josephson effect.

Let us denote the free energy of the specimen as $F$, its minimum value at equilibrium without currents as $-\Gamma$ ($\Gamma > 0$), and the complex order parameter as $\psi(x)$ where $x$ is the position coordinate along the length. The order parameter amplitude when no currents are present is $|\psi_0|$. We assume $\psi(x)$ to be uniform along the cross-section of area $S$, and neglect the effect of magnetic fields arising from the distribution of supercurrents. When expressed in terms of the normalized function $f(x)$=$\psi(x)/|\psi_0|$, the differential equation determining the order parameter is\cite{tinkham,timm}: $\xi^2f''(x) + f(x) -|f(x)|^2f(x)=0$. The solution is given by $f(x)=a_ke^{ikx}$, where the amplitude depends on the wavenumber $k$ as: $a_k$=$\sqrt{1-k^2\xi^2}$. Since $a_k$ is real, allowed values of $k$ need to satisfy $0\!\leq \mid\!k\!\mid\leq\!1/\xi$. The free energy is calculated\cite{supp} to be $F$=$-\Gamma a_k^4$. Using the standard method of Ginzburg-Landau theory, the supercurrent is given by\cite{supp}

\begin{equation}
i_s = i_0~k\xi(1-k^2\xi^2)
\end{equation}

\noindent In the above expression, $i_0$=$\frac{\hbar e}{m_e\xi}|\psi_0|^2S$.  The electrical work done by a voltage $v$ per unit time determines the rate of change of free energy: $\frac{dF}{dt}=i_sv$. This leads to the following result\cite{supp}.

\begin{equation}
v=\frac{\hbar}{2e}\frac{dk}{dt}L
\end{equation}

\noindent This expression provides the necessary insight that a finite voltage in the intermediate state results from dynamic fluctuations of $k$. It implies that both the supercurrent $i_s$ and the superfluid density (given by $|\psi_0|^2a_k^2$) are time-varying quantities when a finite electric field is present.

\begin{figure}
\begin{center}
\includegraphics[width=76mm]{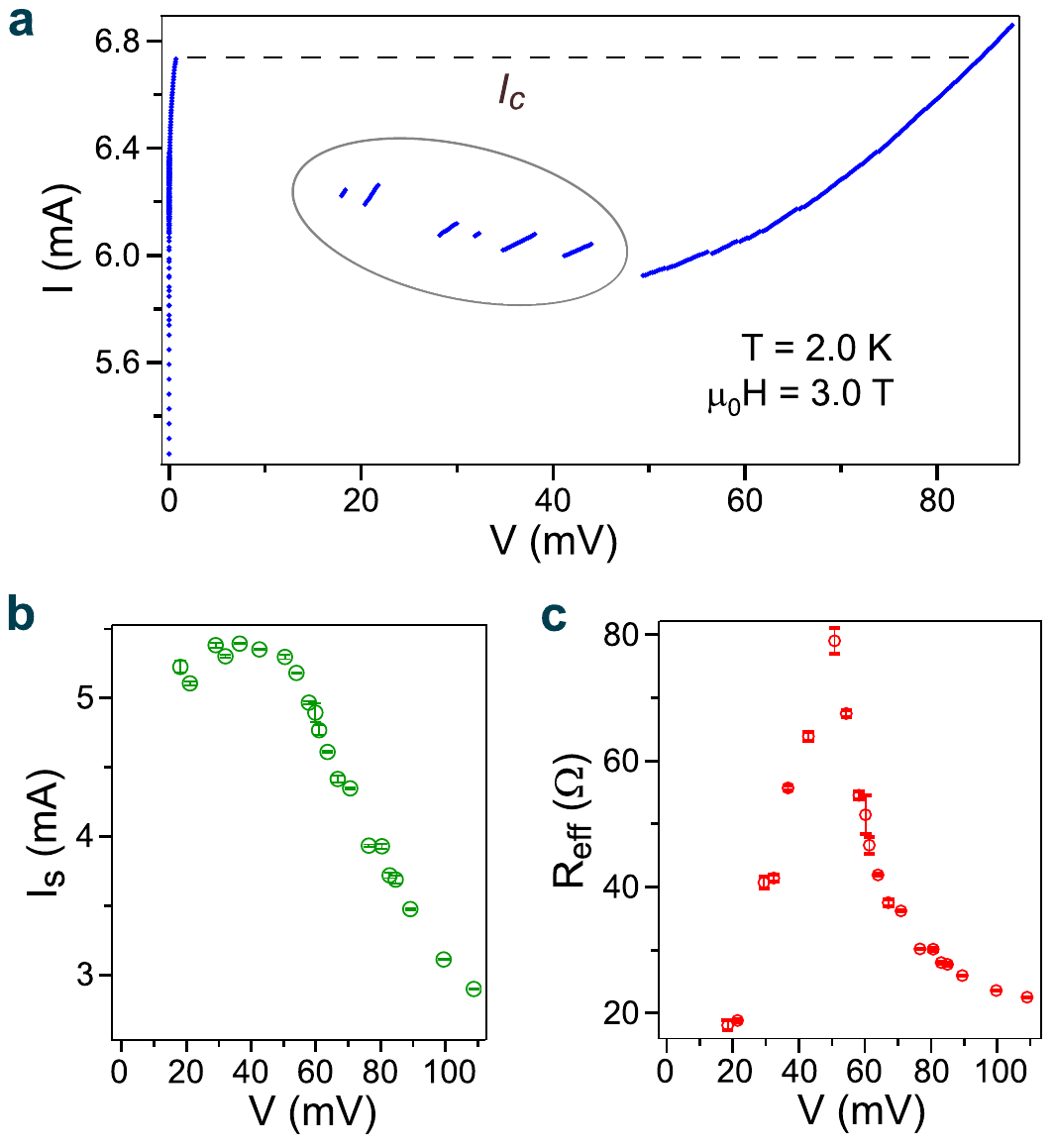}
\caption{\textbf{(a)} $I$-$V$ curve obtained by increasing the source dc voltage $U_0$, with a resistance $R_L$ of 6 $\Omega$ in series with the superconducting film. Very sharp linear segments are seen, with considerable separation along the $V$-axis. \textbf{(b,c)} $I_S$ and $R_{eff}$, estimated from linear fits in different regions of the $I$-$V$ curve, are plotted as a function of $V$.}
\end{center}
\end{figure}

The emergence of the intermediate state (IS) can be understood in terms of a series of resistive domains that appear along the length of the superconducting film (Fig. 3). The sum total of dissipative processes taking place within these domains contribute towards the estimated value of $R_{eff}$. Sudden changes in $R_{eff}$, manifested as jumps from one linear segment to another in the $I$-$V$ plot (as in Figs. 1c and 2a), can be attributed to the formation of new domains. Let us take the $j^{th}$ domain. $i^*_j(t)$ and $i'_j(t)$ are respectively the supercurrent and dissipative current flowing through it, both being functions of time. Our measurements are sensitive to the time-averaged values which satisfy the current conservation criterion $\left\langle i^*_j \right\rangle+\left\langle i'_j \right\rangle=I$. If $r_j$ is the resistance arising from dissipative charge carriers, then the dc voltage drop $v_j$ across this domain can be written as $v_j=r_j\left\langle i'_j \right\rangle$. It then follows that $V=\Sigma v_j$, $R_{eff}=\Sigma r_j$ and $I_S=\Sigma r_j\left\langle i^*_j \right\rangle/R_{eff}$ (where the summation runs over all IS domains). The supercurrent component $I_S$ measured in experiments is therefore to be understood in the sense of an weighted average over the entire sample, with greater weight being assigned to domains having greater values of resistance.

We will now discuss about an estimate of the typical length $\lambda_{IS}$ of an intermediate state domain. If we take the number density of dissipative carriers to be $n$ and the mobility to be $\mu$, then the resistance $r$ of an IS domain should typically scale as $r\sim \frac{\lambda_{IS}}{n\mu}$. We now consider the first appearance of resistance in Fig. 1c, for $V$ = 4.95 mV and $R_{eff}$ = 6.7 $\Omega$. We make a simplifying assumption that $n$ and $\mu$ are the same as the normal state carrier density ($n_0$) and mobility ($\mu_0$) of the Nb film. Then, an approximate estimate of $\lambda_{IS}$ is given by $\lambda_{IS}\approx\frac{R_{eff}}{R_N}L_0$. For the abovementioned values, we obtain $\lambda_{IS}$ = 262 $\mu$m. For a specific value of $R_{eff}$, the total length ($\lambda_T$) of the film in the intermediate state is given by $\lambda_T\approx\frac{R_{eff}}{R_N}L_0$. Both in Figs. 1e and 2c, we find that $R_{eff}$ is of the same order as $R_N$ (14.9 $\Omega$), indicating that the intermediate state appears over a very large fraction of the sample volume. What is quite surprising is that $R_{eff}$ is mostly much larger than $R_N$. We will call this phenomenon as \emph{superresistance}. The fact that $R_{eff}\!>\!R_N$ implies that $\lambda_T\!>\!L_0$, which is unrealistic. This overestimate arises because we have earlier assumed $n\!\approx\!n_0$ and $\mu\!\approx\!\mu_0$. Therefore, either the density or the mobility or both of dissipative carriers responsible for \textit{superresistance} must be smaller than the values in the normal state.

\begin{figure}
\begin{center}
\includegraphics[width=80mm]{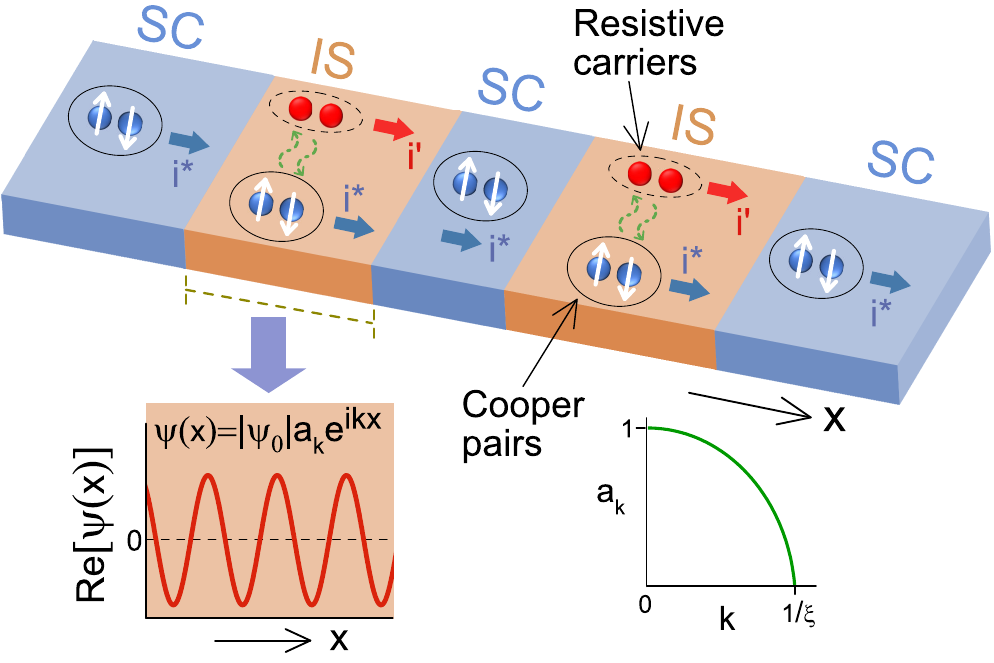}
\caption{When an electric field penetrates into the superconducting film, it is broken up into alternating regions of intermediate state (IS) and superconducting state (SC). The $I$-$V$ relations observed by us can be described as an addition of two current components inside IS domains - a supercurrent ($i^*$) carried by Cooper pairs and a dissipative current ($i'$) carried by resistive carriers. It is not implied that the two channels are spatially separated. Voltage gradients lead to dynamical fluctuations of both the amplitude and phase of the superconducting order parameter.}
\end{center}
\end{figure}

The peculiar nature of the superresistance phenomenon can be appreciated by comparing it to a different effect seen in narrow superconducting wires - phase slip dissipation\cite{tinkham,skocpol}. The latter has its origin in dynamical fluctuations of the order parameter along weak points of quasi-1D superconducting channels, which are called phase slips centres (PSCs). Its key signature is the occurrence of step-like structures in the $I$-$V$ curve, which reveals an excess supercurrent component concomitant with a finite voltage\cite{skocpol,delacour,ivlev,dmitrenko}. On some occassions, similar phenomena in the form of phase slip lines (PSLs) have also been reported in wide superconducting channels (with $w\!\gg\!\xi$) \cite{dmitrenko,sivakov}. In such instances of phase slips, it is generally accepted that the slope increment from one step to another on the $I$-$V$ curve results from normal resistance appearing along a segment of length twice the electric field penetration length ($l_E$)\cite{sivakov}. This results in estimates of $R_{eff}$ based on the $\frac{dI}{dV}$ slope to be smaller than $R_N$, which sharply contrasts with our observation of $R_{eff}\!>\!R_N$. This distinguishes the occurrence of superressitance in the EFDIS  from the scenario of phase slips.

In conclusion, we have shown that a distinct non-equilibrium intermediate state can be induced in superconductors through the application of small voltages, in which features of superconductivity can coexist with electric fields. Its key signature is the simultaneous propagation of a supercurrent and a resistive current. An important insight from our study is that resistive carriers in the intermediate state exhibit markedly different properties from those in the normal state. Further investigation is needed to uncover the microscopic mechanisms underlying these dissipative processes. Our work shows several promising directions for future research. It can motivate experimental studies of intermediate states driven by order parameter dynamics across a broad range of superconducting materials. A particularly interesting class of systems is that of strongly correlated superconductors\cite{keimer}, given their rich phase diagrams and potential for hosting novel forms of quantum matter. The presence of electric fields may compel non-superconducting carriers to participate in charge transport, offering new insights into electronic phenomena that are typically obscured by superconductivity below the critical temperature $T_c$. Another avenue opened by this work is the investigation of structure formation phenomena. It is known that driving systems far from equilibrium may give rise to large-scale dynamical dissipative structures - a topic of considerable interest in fields such as physical chemistry and hydrodynamics\cite{prigogine,glansdorff}. Our observation that the typical lengsthcale of electric-field-induced domains reaches few hundred micrometres suggests that there are intriguing aspects of nonequilibrium structures in the superconducting electronic system.  This merits further attention as it provides a solid-state platform to investigate different topics dealing with the dynamics and dissipative phases of quantum systems \cite{young,dyke,diehl,yamamoto}. Our findings highlight a rich and largely unexplored interface between superconductivity and nonequilibrium statistical mechanics, offering opportunities for future research.

\section{Acknowledgments}
This project was partially funded by a France 2030 funding (PhOM - Graduate School Physique, grant ANR-11-IDEX-0003).



\begin{flushleft}

\newpage

\widetext
\newpage
\begin{center}
\textbf{\large Supplementary Information}
\end{center}
\setcounter{equation}{0}
\setcounter{figure}{0}
\setcounter{table}{0}
\setcounter{page}{1}
\makeatletter
\renewcommand{\theequation}{S\arabic{equation}}
\renewcommand{\thefigure}{S\arabic{figure}}
\renewcommand{\bibnumfmt}[1]{[R#1]}
\renewcommand{\citenumfont}[1]{R#1}

\bigskip

\begin{flushleft}

\bigskip

\textbf{1. Sample fabrication and characterization}

\bigskip

The samples were prepared using commercially available Si wafers with a 500 nm thick layer of insulating SiO$_2$ as the substrate. Devices in four-probe geometry were patterned by designing a mask on polymer resist using electron-beam lithography. Then, Nb was deposited by electron-beam-induced evaporation and the lift-off process was completed by removing the resist layers with acetone. Electrical contacts were established on the Nb samples using a wedge bonder. The cryostat of a Quantum Design Physical Property Measurement System (PPMS) was used for low temperature measurements. A Keithley Model 6487 was used as voltage source. Keithley DMM6500 multimeters were used for voltage measurements. An image of the sample discussed in the main text is shown in Fig. S1.

\bigskip

The upper critical magnetic field ($H_{c2}$) is determined from magnetotransport measurements and the Ginzburg-Landau coherence length ($\xi$) is estimated using the expression $\xi$ = [$\frac{\Phi_0}{2\pi H_{c2}\left(0\right)}$]$^\frac{1}{2}$, where $H_{c2}\left(0\right)$ = $0.69T_c\frac{dH_{c2}}{dT}|_{T=T_c}$ and $\Phi_0$ is the flux quantum\cite{helfand,werthamer}. For the abovementioned sample, we obtain $\xi$ = 10.1 nm. Evaporated films are usually superconductors in the dirty limit, since the mean free path ($l$) is only a few nanometres and smaller than the coherence length ($\xi_0$) of pure crystalline Nb. For dirty superconductors\cite{tinkham2}, $\xi\approx0.85(\xi_0l)^\frac{1}{2}$. Since $\xi_0\approx$ 38 nm for Nb, we estimate $l\approx$ 3.7 nm.

\bigskip

\bigskip

\bigskip

\bigskip

\textbf{2. Complete set of data for determining current-voltage characteristics}

\bigskip

The data for the $I$-$V$ plot in Fig. 1c of the main article was obtained by continuously varying the source voltage $U_0$ (in the circuit of Fig. 1a). We show in Fig. S2 the plots of $I$ and $V$ as a function of $U_0$.

\bigskip

In Fig. S3, we show the variations of $V$ and $I$ as a function of $U_0$ corresponding to Fig. 2a of the main text. The measurements were done by maintaining the sample at 2.0 K temperature with a perpendicular applied magnetic field of 3.0 T. $R_L$ used was 6 $\Omega$. As the source voltage was ramped up starting from zero, the entire voltage dropped initially across $R_L$ with the superconductor being in the zero-resistance state. This continued till the critical current was reached, which in this case occured at $U_0$ = 44 mV. The resistive regions initially appeared outside the part probed by the voltage leads. Finite voltage in four-probe configuration appeared first around $U_0$ = 109 mV, signifying the development of a resistive region somewhere within the length of 584 $\mu$m separating the voltage probes. 

\bigskip

\bigskip

\bigskip

\bigskip

\textbf{3. Measurements upon thermal cycling}

\bigskip

As was evident in Fig. 1c of the main text, there is a fair degree of repeatability for the different sets of $I$-$V$ measurement using four different values of $R_L$. These measurements were done at a fixed value of temperature of 8.30 K. Samples measured in experiments always have a finite disorder distribution which may provide pinning sites at different regions for new domains to appear. From the observations in Fig. 1c, pinning due to defects seems to be somewhat relevant in ensuring that the same configuration of domains is reproduced. However, it is also clear that such disorder potentials do not always play a dominant role, since we often find very different results on thermal cycling - by heating the sample above $T_c$ and cooling back down. For measurements done close to $T_c$ at zero magnetic field, this is sometimes the case (though not always). On the other hand, for measurements done at much lower temperatures at finite magnetic fields, the $I$-$V$ curves are quite often very different upon thermal cycling. In Fig. S4, we compare the results of two different measurements (named Run 1 and Run 2) carried out under identical conditions - before and after a thermal cycle. The data of Run 1 is the same as Fig. 2a of the main article.

\bigskip

\bigskip

\bigskip

\bigskip

\textbf{4. The normal-to-superconductor transition}

\bigskip

In this work, we have focused on inducing the superconductor-to-normal transition by increasing the source voltage starting from zero. It is commonplace for superconductors to show a hysteresis in current-biased critical current measurements. We have observed hysteretic features too. The signatures of the intermediate state are different for the superconductor-to-normal and normal-to-superconductor directions of voltage sweep. Linear segments were found to be present in the $I$-$V$ curves for both sweep directions (Fig. S5a). The $I$-$V$ plot for the superconductor-to-normal transition in Fig. S5a is the same as Fig. 2a of the main article. The values of $I_S$ and $R_{eff}$ estimated for the normal-to-superconductor transition are shown in Figs. S5b and S5c respectively.

\bigskip

\bigskip

\bigskip

\bigskip

\textbf{5. Derivation of the expression for voltage across a long superconducting channel}

\bigskip

We will treat the case of a long superconducting film of length $L$ and uniform cross-sectional area $S$ by means of the Ginzburg-Landau theory. This problem concerns the impact of an electric field in inducing a nonequilibrium state in superconductors, which is different from the issue of suppressing superconductivity with an electric field that has been much discussed in recent literature\cite{simoni,zaccone}. Our objective is to derive the expression of voltage drop along the channel when $L$ is extremely large compared to the superconducting coherence length $\xi$.

\bigskip

The complex order parameter is denoted by $\psi(\textbf{r})$, where $\textbf{r}$ is the position coordinate in three dimensions. The free energy ($F$) is expressed\cite{tinkham2,timm2} in terms of the temperature-dependent phenomenological parameters $\alpha$ and $\beta$.

\begin{equation}
F = \int~d^3\textbf{r}\left[ \alpha \!\mid\!\psi(x)\!\mid^2 +~\frac{\beta}{2}\!\mid\!\psi(x)\!\mid^4 +~ \frac{1}{2m^*}\! \left|\left(\frac{\hbar}{i}\nabla-\frac{e^*}{c}\textbf{A}\right)\psi(x)\right|^2 + \frac{\textbf{B}^2}{8\pi} \right]
\end{equation}

$m^*$ and $e^*$ denote the mass and charge of superconducting charge carriers respectively, $\textbf{B}$ is the magnetic field and $\textbf{A}$ is the vector potential. It is well-known that the superconducting phase is described by a negative value of $\alpha$ and positive value of $\beta$ for $T<T_c$. The value of the order parameter amplitude ($\mid\!\psi_0\!\mid$) at equilibrium, when no current or gradient is imposed, is obtained by minimizing the free energy.

\begin{equation}
\mid\!\psi_0\!\mid = \sqrt{-\frac{\alpha}{\beta}}
\end{equation}

\bigskip

The corresponding value of free energy ($F_0$) is:

\begin{equation}
F_0 = - \Gamma = - \frac{\alpha^2}{2\beta} SL
\end{equation}

\bigskip

The Ginzburg-Landau differential equation\cite{tinkham2} describes the order parameter $\psi(\textbf{r})$ when magnetic fields, currents or gradients are present.

\begin{equation}
\frac{1}{2m^*}\!\left(\frac{\hbar}{i}\nabla - \frac{e^*}{c}\textbf{A}\!\right)^2\!\psi + \alpha\psi~+ \beta\!\mid\!\psi\!\mid^2\!\psi = 0
\end{equation}

\bigskip

The supercurrent density $\mathbf{j_s}$ is given by\cite{tinkham2}:
\begin{equation}
\mathbf{j_{s}} = -i\frac{e^*\hbar}{2m^*}\left(\psi^*\nabla\psi-\psi\nabla\psi^* \right) - \frac{e^{*2}}{m^*c}\!\mid\!\psi\!\mid^2\!\textbf{A}
\end{equation}

\bigskip

We will derive the expressions of supercurrent and voltage using the approximation that magnetic fields are sufficiently small and $\textbf{A}$ can be neglected. We also assume that the order parameter is uniform along the cross-section. This makes it a simplified one-dimensional problem, and we will study the variations of the order parameter along the length only. The position coordinate is now represented as $x$. Eq. S4 can now be re-written in terms of the normalized function $f(x)=\psi(x)/\!\mid\!\psi_0\!\mid$ and the coherence length $\xi$=$\sqrt{-\frac{\hbar^2}{2m^*\alpha}}$.

\begin{equation}
\xi^2\frac{d^2f}{dx^2} + f(x) - \mid\!f(x)\!\mid^2\!f(x) = 0
\end{equation}

\bigskip

The expressions for the dc and ac Josephson effect in a short superconducting weak link can be derived\cite{timm2} starting from Eqs. S6 and S5, using the limiting condition $L<<\xi$. In this  case, the ratio $\frac{\xi}{L}>>1$ and only the first term in Eq. S6 is important. We are interested in the more general case of a long superconducting channel ($L>>\xi$), for which Eq. S6 has to be solved in its entirety. The solution is:

\begin{equation}
f(x)=a_ke^{ikx}
\end{equation}
\begin{equation}
a_k=\sqrt{1-k^2\xi^2}
\end{equation}

$k$ is a wavevector determining the phase of the order parameter $f(x)$. Due to the non-linear term $\mid\!f(x)\!\mid^2\!f(x)$ in Eq. S6, the amplitude $a_k$ has a $k$-dependence as well. Since $a_k$ is real, allowed values of $k$ need to satisfy $0\!\leq \mid\!k\!\mid\leq\!1/\xi$.

\bigskip

The supercurrent density $j_s$, estimated from Eq. S5 and assuming $\textbf{A}$ is negligible, is:
\begin{equation}
j_{s}=\frac{e^*\hbar}{m^*\xi}\mid\!\psi_0\!\mid^2\!k\xi(1-k^2\xi^2)
\end{equation}

\bigskip

The net supercurrent $i_s$=$j_sS$. The free energy (Eq. S1) corresponding to the solution of $f(x)$ in Eq. S7 is estimated to be
\begin{equation}
F=-\Gamma a_k^4
\end{equation}

\bigskip

The rate of change of free energy is given by the condition: $\frac{dF}{dt}=i_sv$, where $v$ is the voltage drop over the distance $L$. This leads to the expression
\begin{equation}
v=\frac{\hbar}{2e}\frac{dk}{dt}L
\end{equation}

\bigskip

In deriving the above expression, we have set $e^*$=2e since the charge of a Cooper pair is twice that of a single electron. According to Eq. S11, the electric field ($v/L$) in a long superconducting channel is given by $\frac{\hbar}{2e}\frac{dk}{dt}$. Since the magnetic vector potential $\textbf{A}$ has been neglected in our analysis, the validity of these results is limited to small values of current and narrow channels. Further corrections need to be made for wide channels carrying larger currents, taking into account the effect of magnetic fields induced by the supercurrent. These calculations are much more complex and would require further work in the future. Despite these limitations, our analysis provides the essential physical insight that finite electric fields inside superconductors are related to a time-varying $k$, which implies fluctuations of both the order parameter phase and amplitude.

\end{flushleft}

\end{flushleft}

\makeatletter
\renewcommand\@biblabel[1]{[R#1]}
\makeatother

\renewcommand{\thefigure}{S\arabic{figure}}

\begin{figure}
\begin{center}
\includegraphics[width=100mm]{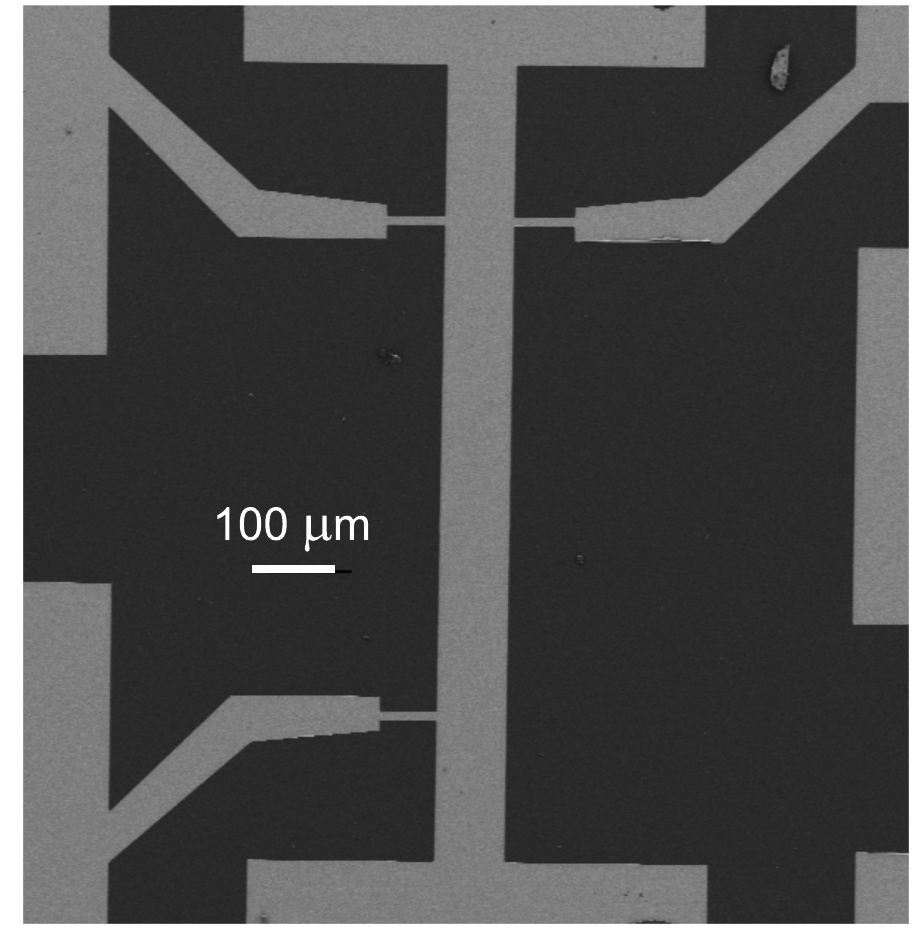}
\caption{\textbf{Scanning electon micrscope (SEM) image of the Nb film described in the main article.}}
\end{center}
\end{figure}

\begin{figure}
\begin{center}
\includegraphics[width=168mm]{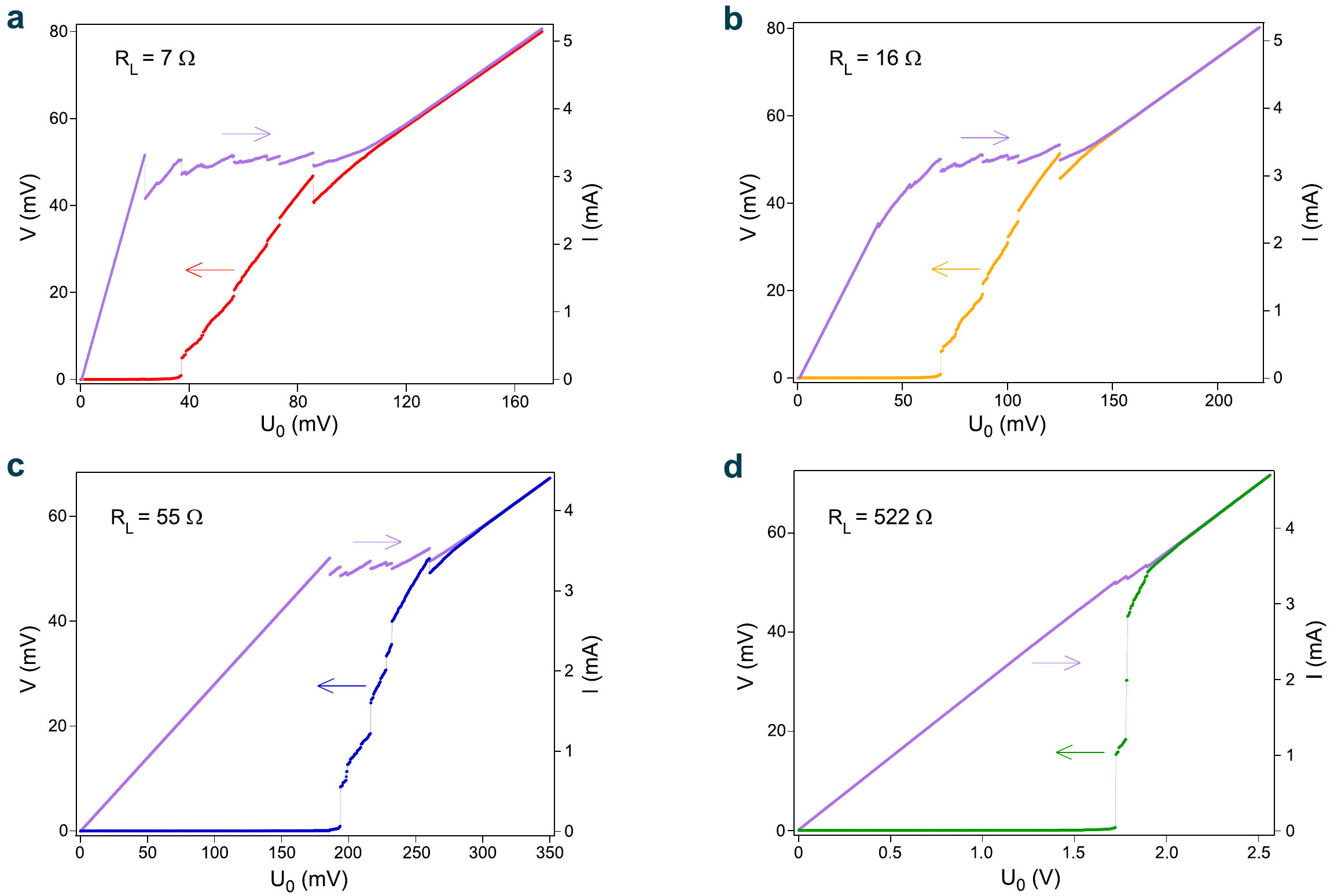}
\caption{\textbf{The plots of voltage across the superconductor ($V$) and current ($I$) as a function of the source voltage $U_0$.} \textbf{(a)-(d)} The results are shown for four different values of $R_L$ in the circuit of Fig. 1a of the main text. The $I$-$V$ plot in Fig. 1c of the main text is generated from this set of data.}
\end{center}
\end{figure}

\begin{figure}
\begin{center}
\includegraphics[width=168mm]{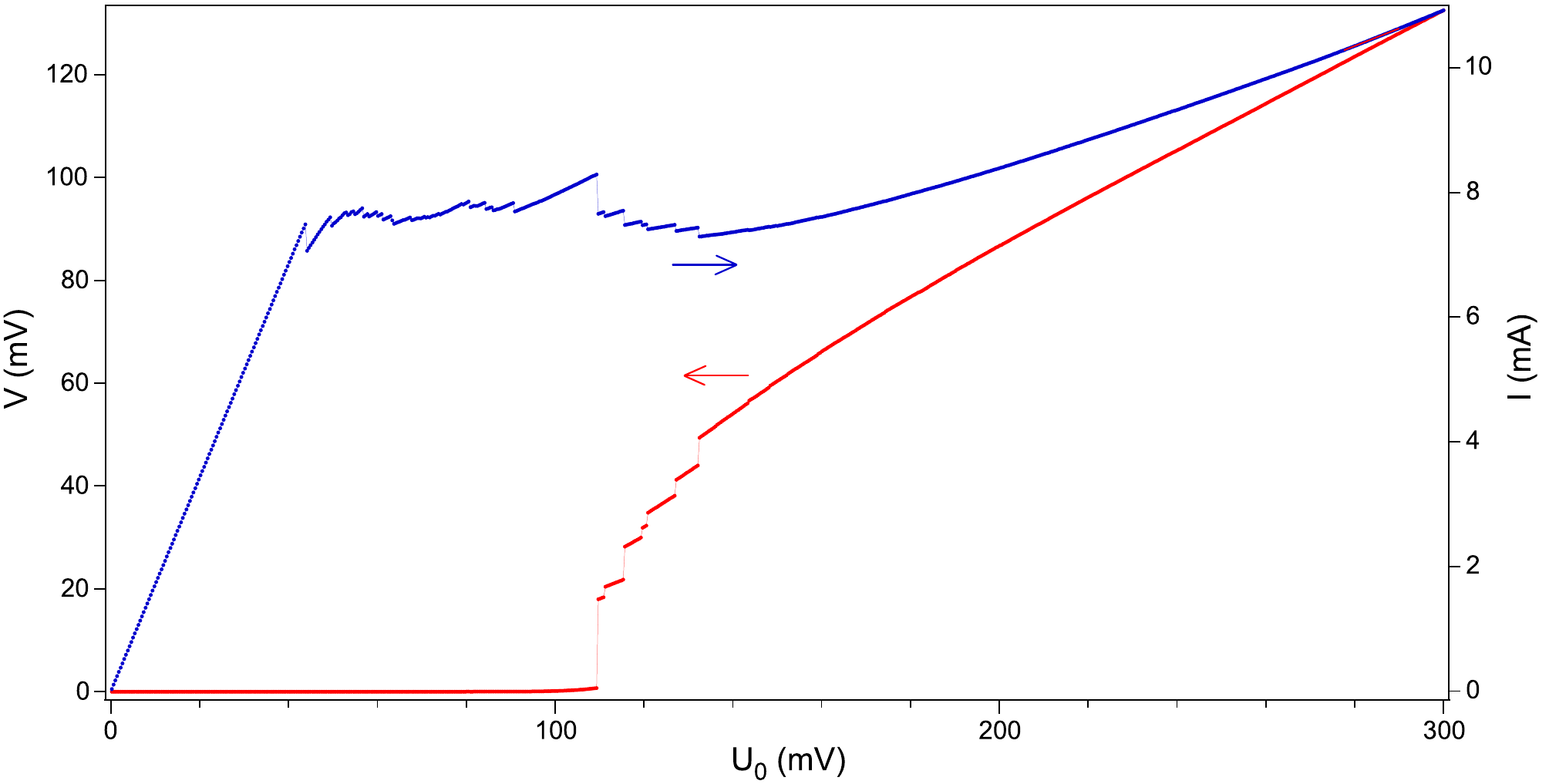}
\caption{\textbf{The plots of ($V$) and ($I$) as a function of $U_0$ corresponding to Fig. 2a of the main article.} These measurements were done by maintaining the sample at 2.0 K temperature with a perpendicular applied magnetic field of 3.0 T. $R_L$ used was 6 $\Omega$.}
\end{center}
\end{figure}

\begin{figure}
\begin{center}
\includegraphics[width=120mm]{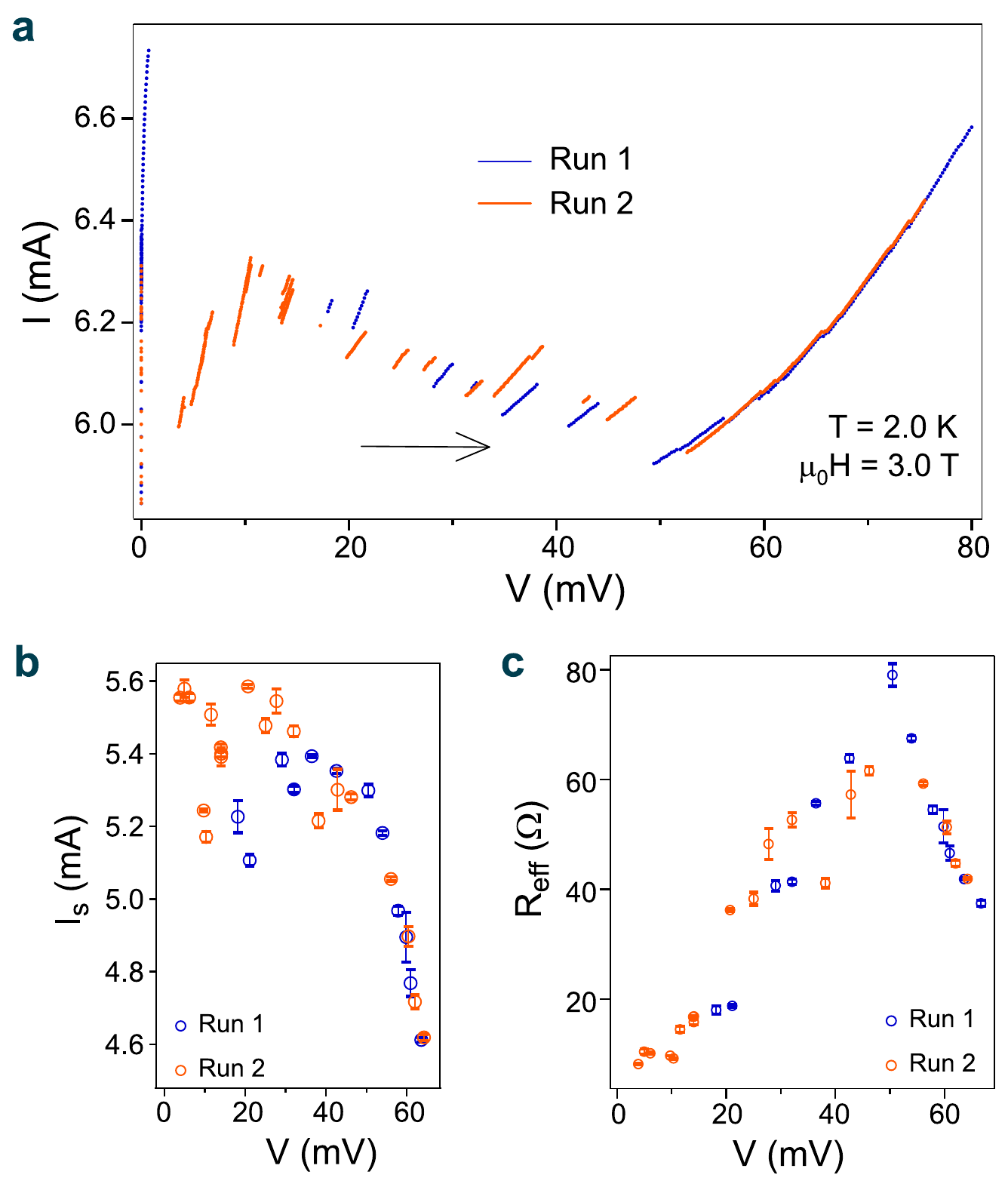}
\caption{\textbf{Differences in $I$-$V$ plots upon thermal cycling of the sample.} \textbf{(a)} Measurements done at $T$ = 2.0 K and $\mu_0$H = 3.0 T using $R_L$ of 6 $\Omega$. \textbf{(b,c)} $I_S$ and $R_{eff}$ estimated for the two different measurements.}
\end{center}
\end{figure}

\begin{figure}
\begin{center}
\includegraphics[width=140mm]{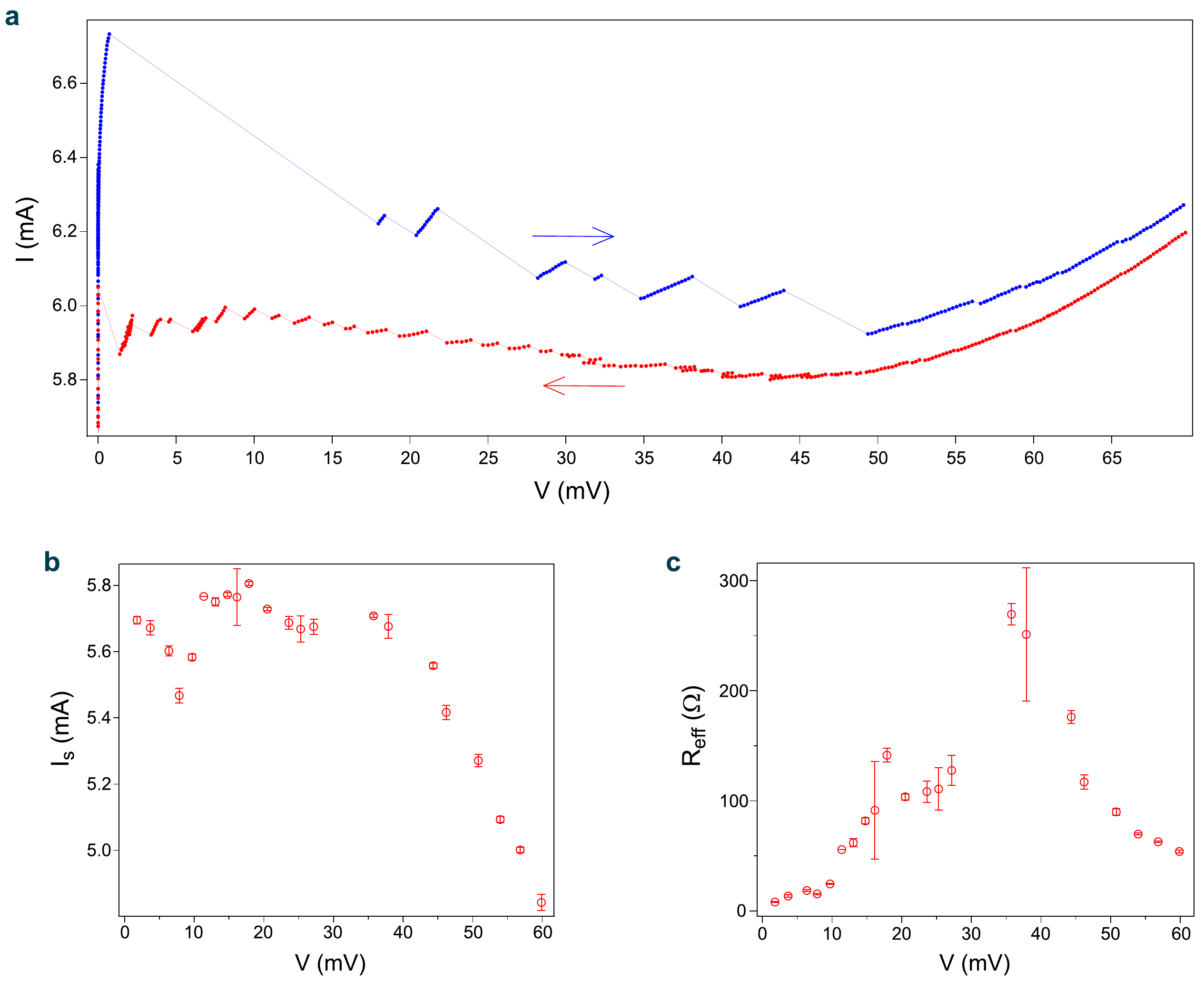}
\caption{\textbf{$I$-$V$ plots for both directions of source voltage sweep.} \textbf{(a)} Linear segments are present for both the superconductor-to-normal (blue) and normal-to-superconductor (red) transitions. The measurements were done at $T$ = 2.0 K and with $\mu_0$H = 3.0 T using $R_L$ of 6 $\Omega$. \textbf{(b,c)} $I_s$ and $R_{eff}$ for the normal-to-superconductor transition.}
\end{center}
\end{figure}

\end{document}